\documentclass{article}
\usepackage{emulateapj,timesfonts}
\usepackage{float}
\usepackage{pstricks}
   \def\apj#1#2#3#4{\par #4 19#3, {\em Ap. J.,\/} {\bf #1}, #2 }
   
   \def\apjs#1#2#3#4{\par #4 19#3, {\em Ap. J. (Supplement Series),\/} {\bf #1}, #2 }
   \def\aj#1#2#3#4{\par #4 19#3, {\em Astr. J.,\/} {\bf #1}, #2}
   \def\pasj#1#2#3#4{\par #4 19#3, {\em PASJ,\/} {\bf #1},#2 }

   \def\ogip#1#2#3{\par #3 #2 version, MPE / OGIP Calibration Memo CAL/ROS/#1}

   \def\aa#1#2#3#4{\par #4 19#3, {\em A\&A,\/} {\bf #1}, #2 }
   
   \def\aas#1#2#3#4{\par #4 19#3, {\em A\&A (Supplement Series),\/} {\bf #1}, #2 }
   
   \def\mnras#1#2#3#4{\par #4 19#3, {\em M.N.R.A.S.,\/} {\bf #1}, #2 }

   \def\apjl#1#2#3#4{\par #4 19#3, {\em Ap. J. (Letters).,\/} {\bf #1}, #2 }

   \def\astroph#1#2#3{\par #3 19#2, {\em Preprint Astro-ph} No. #1}
   \def\BIB {\par}

\def\einstein{{\it Einstein}}

\def\hea4{{\it HEAO~A4}}
\def\heaoa2{{\it HEAO~A2}}
\def\heao1{{\it HEAO~1}}

\def\amin{$^\prime$}
\def\asec{$^{\prime\prime}$}

\def\eg{{\it e.g.}~}

\def\deg{$^{\circ}$}

\def\h0{$H_{\rm o}=50$~km~s$^{-1}$~Mpc$^{-1}$}
\def\q0{$q_{\rm o}$}

\def\msun     {$M_{\odot}$}
\def\lsun     {$L_{\odot}$}

\def\etal    {{ et~al.}~}

\def\ergssec   {~ergs~sec$^{-1}$}
\def\ergss   {~ergs~s$^{-1}$}
\def\ergscmsec   {~ergs~cm$^{-2}$~sec$^{-1}$}

\def\cms3  {~{cm$^{-3}$}}
\def\nhl{~{$N_{\rm H}$}}

\def\nd{{\mbox{0.5--2.0~keV}}}
\def\nsd{{\mbox{0.5--0.9~keV}}}
\def\nhd{{\mbox{0.9--2.0~keV}}}

\def\nhl{N$_{\rm H}$}
\def\kte{{\it k}T$_{\rm e}$}

\def\tild{^{^{\hspace*{-5pt}\sim}}}
\begin{document}
\submitted{Accepted for publication in ApJ}

\title{{Stellar Metallicities and SN~Ia Rates in the Early-type Galaxy NGC5846}\\
{from ROSAT and ASCA Observations}}

\author{A.~Finoguenov$^{1,2}$, C.~Jones$^1$, W.~Forman$^1$ and L.~David$^1$}
\affil{{$^1$ Harvard-Smithsonian Center for Astrophysics, 60 Garden st., MS 2,
    Cambridge, MA 02138, USA}\\
{$^2$ Space Research Institute, Profsoyuznaya 84/32 117810 Moscow, Russia}}
\authoremail{alexis@hea.iki.rssi.ru}

\begin{abstract}

In this paper we analyze the diffuse X-ray coronae surrounding the
elliptical galaxy NGC5846, combining measurements from two
observatories, ROSAT and ASCA. We map the gas temperature distribution
and find a central cool region within an approximately isothermal gas
halo extending to a radius of about 50 kpc, and evidence for a
temperature decrease at larger radii. With a radially falling
temperature profile, the total mass converges to $9.6\pm1.0 \times
10^{12}${\msun} at $\sim$230~kpc radius. This corresponds to a total
mass to blue light ratio of $53\pm5$ \msun/\lsun. As in other early
type galaxies, the gas mass is only a few percent of the total mass.

Using the spectroscopic measurements, we also derive radial distributions
for the heavy elements silicon and iron and find that the abundances of both
decrease with galaxy radius. The mass ratio of Si to Fe lies between the
theoretical predictions for element production in SN~Ia and SN~II,
suggesting an important role for SN~Ia, as well as SN~II, for gas enrichment
in ellipticals. Using the SN~Ia yield of Si, we set an upper limit of
$0.012h_{50}^2$ SNU for the SN~Ia rate at radii $>50$~kpc, which is
independent of possible uncertainties in the iron L-shell modeling. We
compare our observations with the theoretical predictions for the chemical
evolution of ellipticals, taken from Matteucci \& Gibson (1995). We conclude
that the metal content in stars, if explained by the star formation
duration, requires a significant decline in the duration of star formation
with galaxy radius, ranging from $\sim1$~Gyr at the center to $\sim0.01$~Gyr
at 100~kpc radius.  Alternatively, the decline in metallicity with galaxy
radius may be caused by a similar drop with radius in the efficiency of star
formation. Based on the Si and Fe measurements presented in this paper, we
conclude that the latter scenario is preferred, unless a dependence of the
SN Ia rate on stellar metallicity is invoked.

\end{abstract}


\section{Introduction}

X-ray observations have shown that early type galaxies are gas rich
systems with gas masses up to several 10$^{10}$ $M_{\odot}$ and
temperatures around $10^{7}$~K (e.g. Forman \etal 1979; Forman, Jones
\& Tucker 1985). The heavy element abundances in these hot coronae and
their radial dependencies should reflect input to the corona from
evolving stars and supernovae.  Since the flow time of the gas in the
corona is long (billions of years), spatially resolved X-ray
spectroscopy of the gas can provide constraints on abundances in the
stellar component and the rate of supernovae (Loewenstein and Mathews
1991, hereafter LM). Measuring the elemental abundances in the
extended coronae of galaxies is complimentary to measuring abundances
from the integrated optical light of stars, which is characteristic
of the central region in a galaxy.

With the spectral resolution of the ROSAT PSPC, only the iron abundance
could be reliably measured in the X-ray coronae of early-type galaxies (\eg
Forman \etal 1993, Davis and White 1996). With the higher spectral
resolution of ASCA, we can now separate the spectral features of several
heavy elements, which allows us to distinguish between the impact of stellar
mass loss and SN Ia input.

To explore the abundances in the hot gas around early-type galaxies, we
analyzed both ASCA and ROSAT observations of NGC5846, the brightest galaxy
($L_B=1.82\times10^{11}$) in the LGG393 group (Garcia 1993 and references
therein). The NGC5846 group is also known as cV50 and CfA 150. NGC5846 is
classified as an S0$_1$(0) galaxy by Sandage and Tammann (1987), although it
is more likely an elliptical of E0 type, as classified in RC2 (\eg\ Wrobel
\& Heeschen 1991).  The galaxy's velocity relative to the Local Group is
1674 km s$^{-1}$, corresponding to a distance of 33.5 Mpc (\h0) which
implies a scale of $\sim10$ (9.74) kpc per arcminute.  An optical study of
the NGC5846 group yielded a velocity dispersion of $\sigma=381$ km s$^{-1}$
and M/L=163 in solar units (Haynes \& Giovanelli 1991).  For NGC5846, the
internal velocity dispersion is $\sigma=230$ km s$^{-1}$ (\eg\ Fisher,
Illingworth \& Franx 1995).  As one of only a few nearby, bright elliptical
galaxies, NGC5846 is also bright in X-rays, so that the X-ray properties of
the corona and the underlying mass distribution can be studied with the
ROSAT and ASCA observatories.  The total X-ray luminosity of NGC5846 inside
a 10\amin\ radius is $7\times10^{41}$ \ergssec, in the 0.2--2.0 keV band.
NGC5846 was previously studied in X-rays by Biermann \etal (1989) using
{\einstein} observations, who assuming an isothermal coronae, derived a
gravitating mass inside a 120 kpc radius of $7\pm2\times10^{12}$\msun\ and a
total gas mass of $8.5\times10^{10}$\msun.

In the sections which follow, we describe the ROSAT and ASCA observations of
NGC5846.  Section 2 discusses the analysis of both the imaging and
spectroscopic data.  Section 3 briefly describes our results for the spiral
galaxy NGC5850, which is detected in both the ROSAT and ASCA images. In
section 4 we discuss the mass determination around NGC5846 and our heavy
element abundance measurements. Finally, we described in the Appendix our
approach to analyzing extended sources when observed with the broad, energy
dependent PSF of the ASCA X-ray telescopes.

\section{ ROSAT and ASCA Observations and Analysis}

For imaging analysis and to map the temperature distribution of the emission
around NGC5846, we used ROSAT PSPC observations carried out during 25 July
-- 8 August 1992 and a second observation performed during 18 January 1993.
We used the software described in Snowden \etal (1994) and references
therein to determine the ``good time'' intervals (GTI's), to estimate the
non-X-ray background, and to construct exposure maps for flat fielding.

We observed NGC5846 with ASCA during 7--8 February 1994 for a total of 30
ksec. In our analysis, we used data from the SIS0 and SIS1
detectors. The four-CCD SIS mode, which we used, provides a FOV of
20\amin$\times$20\amin\ with an energy resolution of $\sim$75~eV at 1.5 keV
(preflight value). A detailed description of the ASCA observatory, as well as
the SIS detectors, can be found in Tanaka, Inoue \& Holt (1994) and Burke
\etal (1991). Standard data screening was carried out with FTOOLS version
3.6.  In our spectral analysis, we account for the effects of the broad ASCA
PSF, as described in {\it Appendix} of this paper, and include the
geometrical projection effect of the three-dimensional distribution of
emitting gas. GIS data were not used in our analysis, due to their lower
energy resolution.

Finally, we compare the results of a three-dimensional approach to modeling
the ROSAT and ASCA observations. In such an approach, where all the derived
spectra are fitted simultaneously and where correlations between different
regions are high, a simple minimization of the $\chi^2$ term is shown to be
an ill-conditioned task (Press \etal 1992, p.780). Among the regularization
algorithms developed to solve this problem (Press \etal 1992, p.801), we
chose the assumption that a linear function is a good representation of the
radial distribution of the gas temperature and abundance.  The impact of the
regularization is limited to changing the best-fit values only within their
68\% confidence level intervals. Thus, the procedure results in a linear
solution for the parameters of interest in the regions where they could not
be resolved either statistically or spatially. Further details of this
minimization procedure for ROSAT and ASCA observations are described in
Finoguenov and Ponman (1998).  We adapted the XSPEC analysis package to
perform the actual fitting.

\subsection{ Imaging Analysis}
\subsubsection{ ROSAT PSPC }

Since ROSAT has superior spatial resolution compared to ASCA, we used the
ROSAT PSPC images to determine both the X-ray surface brightness and the gas
temperature distribution for NGC5846. For surface brightness analysis, we
used the broad energy band {\nd} (Snowden bands R47). We first generated an
auxiliary map, in which we identified point-sources and sharp emission
features. We used this map to determine the smoothing kernel for R47 and
also for the gas temperature analysis, so that the sharp features would not
be strongly smoothed. In addition to smoothing the background subtracted
image, an identical smoothing was applied to the exposure map. With these
two images, we generated a vignetting corrected, background subtracted image
(counts sec$^{-1}$ cm$^{-2}$ arcmin$^{-2}$) for all regions where the
exposure map exceeded 1\% of the map's maximum. For NGC5846 this procedure
results in smoothing the emission of the core and all point sources with a
gaussian of $\sigma$=30\asec, the inner corona with $\sigma$=1\amin\, and
the outskirts ($>5$\amin) of the galaxy, with $\sigma$=2\amin. We use this
image to determine the center of the X-ray emission.

\bigskip
\centerline{\includegraphics[width=3.25in]{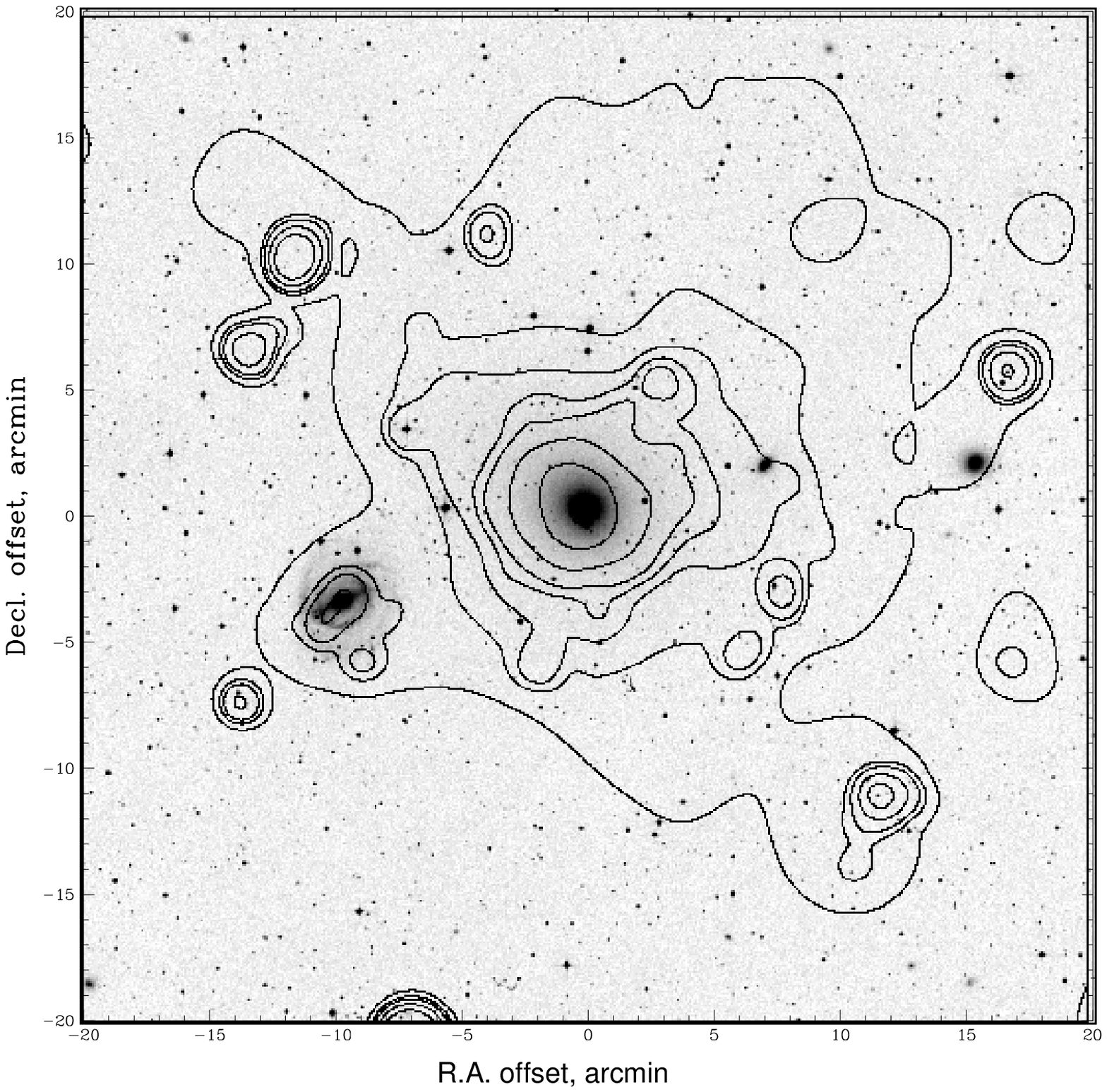}}

\figcaption{Optical image of the NGC5846 field with the
  overlaying adaptively smoothed X-ray intensity contours.  Except for NGC5846,
  the only other galaxy detected in X-rays is NGC5850.  North is towards the 
  top, east is to the left. The center of the figure is 15$^h$ 06$^m$
29.2$^s$; +01\deg\ 36\amin\ 17\asec (Eq. J2000).
\label{opt_imh}}
\medskip

In Figure \ref{opt_imh} we present an optical image of the NGC5846 field
with the ROSAT broadband (R47) contours superposed.  X-ray contours extend
to a radius of 200 kpc, far beyond the detected optical light of NGC5846.
While there are numerous serendipitous sources in the image, in addition to
NGC5846, the only galaxy detected is NGC5850 (type SBb, Sandage \& Tammann
1988), which lies beyond the N5846 group.

\begin{minipage}[H]{8.5cm}
\begin{table}[H]

{\renewcommand{\arraystretch}{1.4} \renewcommand{\tabcolsep}{0.3cm}
\begin{center}
\tabcaption{\centerline{\footnotesize
Radial Profile Fitting of NGC5846${^a}$.}}
\footnotesize

\begin{tabular}{lllll}
\hline
\hline
\vspace{1pc}
{$\begin{array}{c}\mbox{Boundaries}\\
    \mbox{(arcmin)}\end{array}$} & $\beta$ & {$\begin{array}{l}\mbox{$r_a$}\\
                  \mbox{(arcmin)}\end{array}$} &
$\chi^2_r / d.o.f.$\\
\hline
0 -- 12 $^{^b}$ & 0.54$\pm0.06$  & 0.64$_{-0.18}^{+0.34}$  &  \\
0 -- 20 & 0.568$\pm$0.009 & 0.527$\pm0.02$ & 1.55/122  \\
0 -- 20 $^{^c}$ & 0.577$\pm$0.008 & 0.473$\pm0.02$ & 1.41/122 \\
0 -- 20 $^{^d}$ & 0.553$\pm$0.008 & 0.425$\pm0.02$ & 1.45/122 \\
\hline
\end{tabular}
\end{center}
}

{$^a$}{\footnotesize ~ Profile
was derived from ROSAT PSPC data in the {\mbox{0.5--2.0~keV}} band.  Errors
are given at 90\% confidence level for one parameter of interest. See text
for details of analysis.}

{$^b$}{\footnotesize ~ {\einstein} data from Biermann \etal (1989)}

{$^c$}{\footnotesize ~ including observed temperature gradient}

{$^d$}{\footnotesize ~ including observed temperature and abundance
gradients}

\end{table}
\end{minipage}

Using the ROSAT PSPC images, we extracted a radial surface brightness
profile around NGC5846, corrected it by the corresponding exposure map
profiles, and added a 4\% systematic error.  We use a $\beta$-model of the
form \[S(r)=S(0) \; (1+({r / r_a})^2)^{-3\beta+{1\over 2}} \] to characterize
the surface brightness profile.  We use the region 20\amin--40\amin\ for
background estimation. Our results, listed in Table 1, provide a better
determination of $r_a$ and $\beta$, but are consistent with the {\einstein}
analysis (Biermann \etal 1989). The best fit and the surface brightness
profile are presented in Figure \ref{rad_pro}. Since the ROSAT PSPC
countrate is affected by variations in the gas temperature and abundance, we
take into account our determinations of the projected temperature and
abundance in our analysis of the surface brightness profile, which we also
use for gas mass calculations. As shown in Table 1, changes occur primarily
in the derived values of the core radius ($\pm10$\%), with only $\pm2$\%
changes in the $\beta$ index.

\bigskip
\centerline{\includegraphics[width=3.25in]{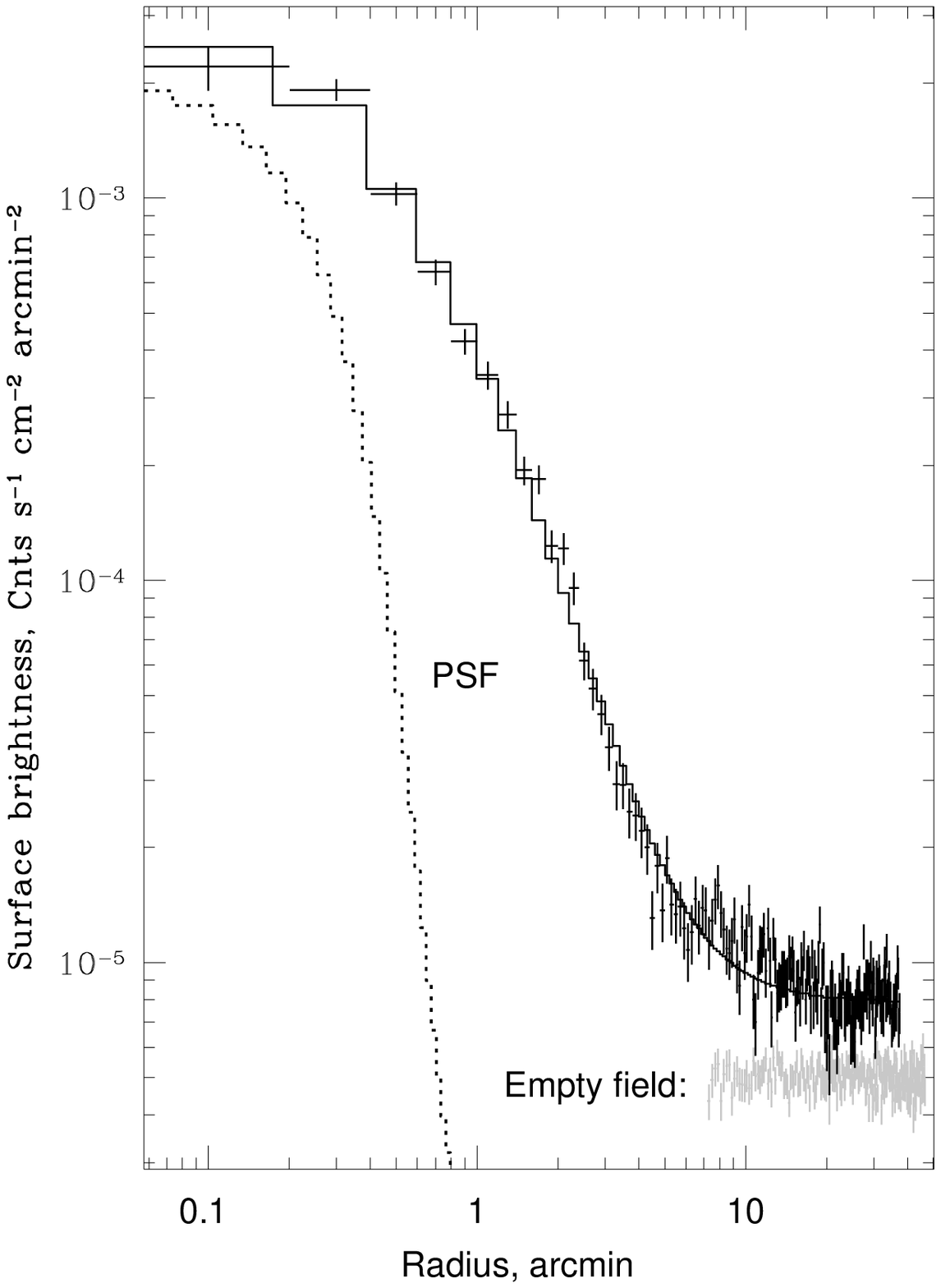}}

\figcaption{The radial surface brightness profile of the X-ray emission from
  NGC5846 derived from the PSPC observations.  The ROSAT PSPC PSF which was
  generated for a point source with the position and spectrum of NGC5846 is
  shown for comparison to the NGC5846 surface brightness profile. The solid
  curve represents a best fit to the NGC5846 profile by a $\beta$-model
  convolved with the PSF. Parameters of the best fit are $r_a=0.43$
  arcminutes and $\beta=0.55$.  Errors are shown at the 1$\sigma$ level.
  Background behavior at large off-axis angles is illustrated by the grey
  points, extracted from an empty field and scaled to the NGC5846 image.
\label{rad_pro}}
\medskip

To analyze the gas temperature distribution, we generated a hardness ratio
map.  We choose two semi-independent energy bands, {\nsd} (R45) and {\nhd}
(R67) (see Finoguenov \etal 1998 for a discussion of the particular choice
of these bands). We used the point source map and minimal smoothing scale
$\sigma=1$\amin\ to avoid small-scale artifacts, which would otherwise arise
from the energy dependence of the ROSAT PSPC PSF (Hasinger \etal 1994).
Diffuse emission from NGC5846 within a radius of 5\amin\ was smoothed with
$\sigma=2$\amin\ and the rest with $\sigma=10$\amin.

\bigskip
\centerline{\includegraphics[width=3.25in]{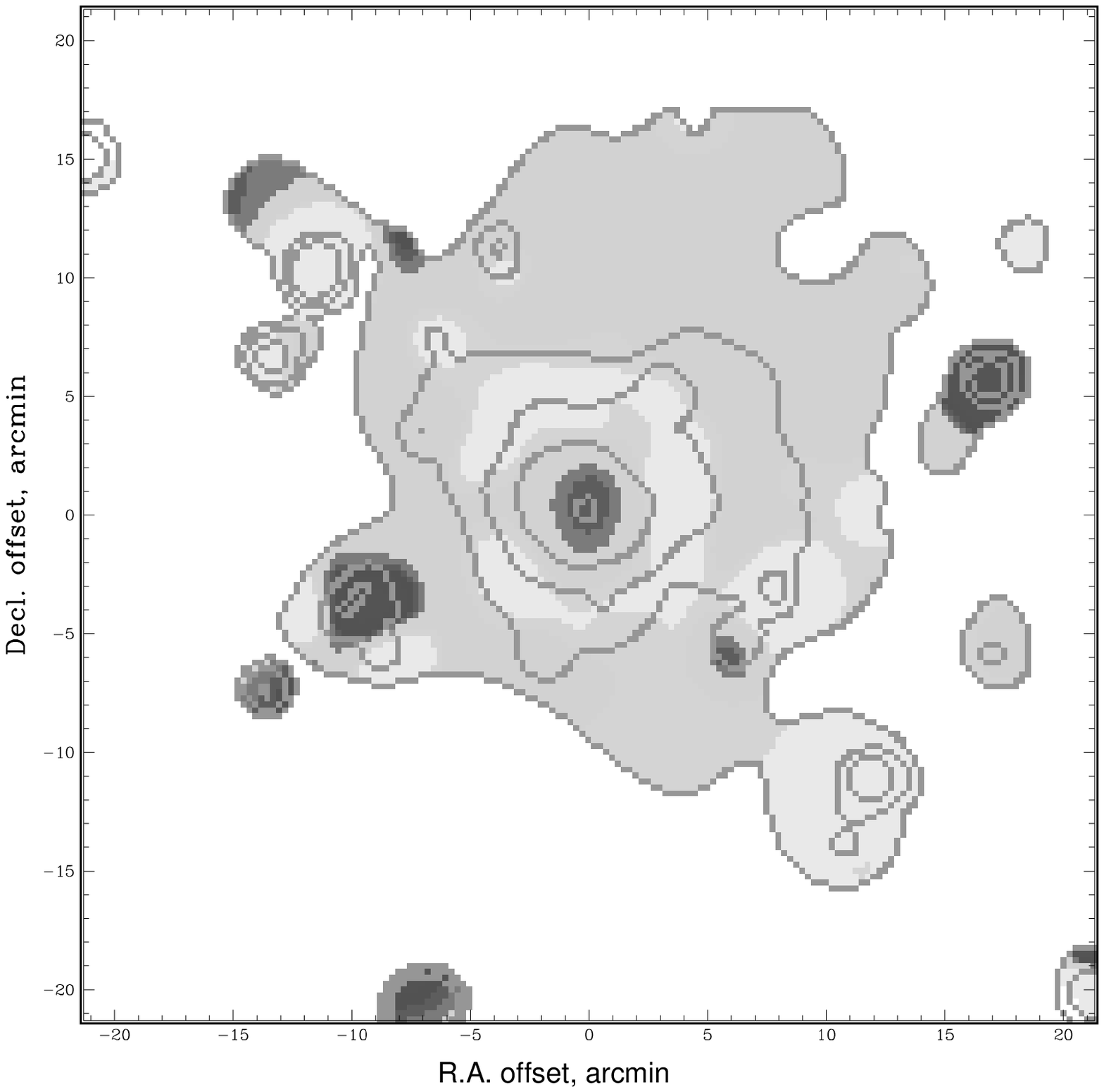}}

\figcaption{The ROSAT hardness ratio R67 to R45 (R67--R45)/(R67+R45)
  overlaid with the X-ray surface brightness contours (R47). Three levels of
grey correspond to different temperatures. Temperatures of $\sim$0.6 keV are
indicated by dark grey, $\sim$0.8 keV by grey, and $\sim$1.0 by light grey.
Variations outside the outer contour of surface brightness are not
statistically significant and are omitted. The center of the figure is
15$^h$ 06$^m$ 29.2$^s$; +01\deg\ 36\amin\ 17\asec (Eq. J2000).
\label{te_map}}
\medskip

In Figure \ref{te_map} we present a hardness ratio image defined as
(R67--R45)/(R67+R45) along with the broad-band X-ray surface brightness
contours.  This hardness ratio provides a robust estimator of the
temperature in the range 0.5--1.5 keV, even if \nhl\ and abundances vary
(Finoguenov \etal 1998).  As Figure \ref{te_map} shows, the X-ray emission
appears cool at the galaxy center, with the temperature increasing smoothly
with radius to $\sim1$~keV at $\sim$3\amin\ from the center.  At radii
beyond 5\amin\, the emission has a cooler temperature (\kte$\sim0.8$~keV).

\subsubsection{ ASCA SIS}

The ASCA SIS imaging capability is limited compared to the ROSAT PSPC, due
to the broader ASCA PSF. Nevertheless, analysis of the ASCA images is important
for several respects, in particular to measure the distribution of harder
($>2$ keV) emission.  To analyze the ASCA observations, we first determine
the alignment between the ROSAT and ASCA images. We use the results of the
ROSAT profile fitting to simulate the ASCA image in the \mbox{0.6--2.0~keV}
range and then find the relative alignment with the true ASCA image. Our
determined ASCA position for the NGC5846 center is
\ 15$^h$ 06$^m$ 29.9$^s$; +01\deg\ 36\amin\ 21\asec\ (RA; Dec., Equinox
J2000) compared with the ROSAT position
\ 15$^h$ 06$^m$ 29.2$^s$; +01\deg\ 36\amin\ 17\asec,
corrected for the ROSAT XRT/detector boresight offset (Briel \etal 1993).
The relative misalignment between ASCA and ROSAT is within the expected
uncertainties in both ASCA (Gotthelf 1996) and ROSAT (Briel \etal 1993)
aspect determination.  For both ASCA and ROSAT, the X-ray center is
consistent with the optical center of the galaxy, \ 15$^h$ 06$^m$ 29.26$^s$;
+01\deg\ 36\amin\ 20.7\asec.

\bigskip
\centerline{\includegraphics[width=3.25in]{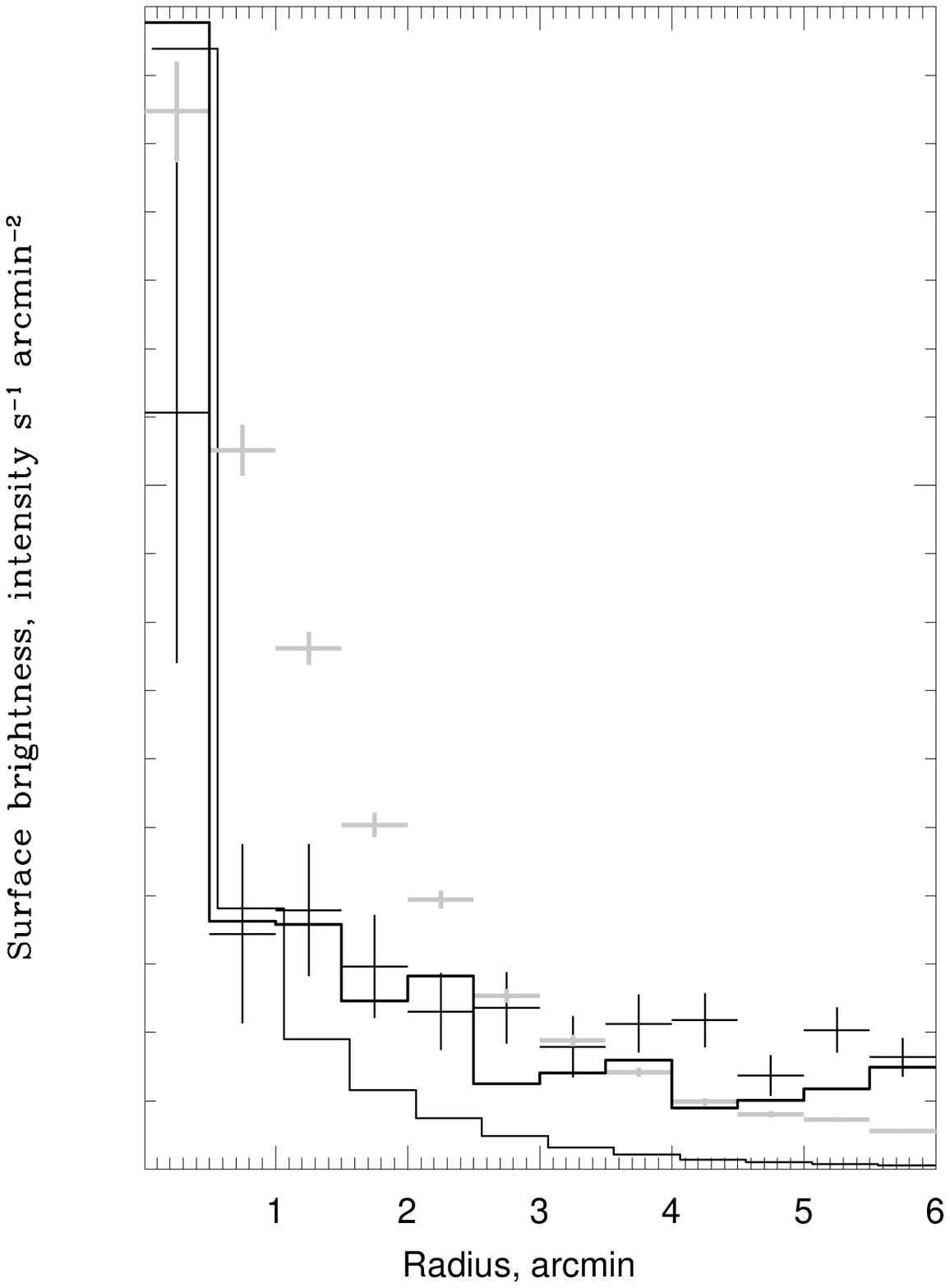}}

\figcaption{Radial surface brightness profile of the NGC5846 X-ray emission
  detected with ASCA (background subtracted). Black crosses are the profile
derived in the 3--6 keV energy band, and grey crosses correspond to the
0.7--2.5 keV band. The thick solid line represents the ASCA PSF plus the
estimated background.  The thin solid line denotes the ASCA PSF.
\label{hard_pro}}
\medskip

The \mbox{0.6--2.0~keV} ASCA image also detects an X-ray source at the
position of NGC5850. We also examined the ASCA image in the harder
\mbox{3--6~keV} band, and found  only a peak at the galaxy center, with no
detected surrounding diffuse emission (see source profile in
Fig.\ref{hard_pro}). For our analysis, we used the 3--6 keV image to
determine the normalization for the SIS blank sky background. Finally, we
note that in analyzing the 0.4--0.6 keV ASCA image, in the outer regions of
the SIS chips, we found a possible additional background component with a
very soft spectrum.

\subsection{ Spectral Analysis of NGC5846} \label{sec:spe}

In the spectral modeling of NGC5846 from both the ROSAT and ASCA
observations, we used a single temperature MEKAL model (Mewe \etal 1985,
Liedahl \etal 1995), modified by absorption, with the column density fixed
to the galactic value of $4\times10^{20}$ cm$^{-2}$ (Stark \etal 1992) and
the source redshift of 0.0058. The solar values for Si and Fe are defined as
3.55e-5 and 4.68e-5 by number, compared to hydrogen (Anders \& Grevesse
1989). Fitting was done over the energy ranges of 0.2--2.0 keV for ROSAT and
0.7--2.2 keV for ASCA. For PSPC spectral analysis elements other then Fe
are fixed to the ASCA values.

Our choice of the MEKAL plasma code leads to systematically lower
temperatures (at the 20\% level), compared to the Raymond-Smith code
(Raymond \etal 1977). Also, as was shown in a study by Matsushita (1998), if
Fe abundance is decoupled from the other heavy elements, differences in the
derived Fe abundances are only 20--30\%. If there is excess absorption at
the galaxy center, only the innermost ROSAT abundance measurements would be
affected by fitting the spectra with fixed galactic absorption. Omitting the
low-energy channels in the analysis of ASCA spectra results in the absence
of sensitivity to both the O abundance level and small variations in
galactic column. Due to the complications of cooling, as well as absorption,
we do not present the abundance results for the central region.

To perform a spectral analysis for both ASCA and ROSAT, an ARF matrix was
calculated, using a 3-dimensional model of the source, based on the surface
brightness profile, derived from ROSAT data and taking the PSF into account
(see Appendix).  Although the count rate in the 0.5--2.0 keV ROSAT PSPC band
is strongly affected by variations in temperature and abundance for values
typical for NGC5846, a problem would arise only if these changes are within
one data bin, chosen for further analysis. Otherwise changes only affect the
normalization, which we remove in our calculation of the scattering matrix,
by normalizing the redistribution of counts on the input image.

To exclude serendipitous sources from the ROSAT spectra of NGC5846, we
omitted a circular region around each detected source. The radius of
the circle was chosen to match the radial dependence of the PSF, with
1\amin\ the smallest radii for on-axis sources.  To generate the ASCA
spectra for NGC5846, we subtract the contribution from background
sources by normalizing the ``blank sky'' reference spectrum to our
ASCA image in the 3--6~keV band.

\bigskip
\centerline{\includegraphics[width=3.25in]{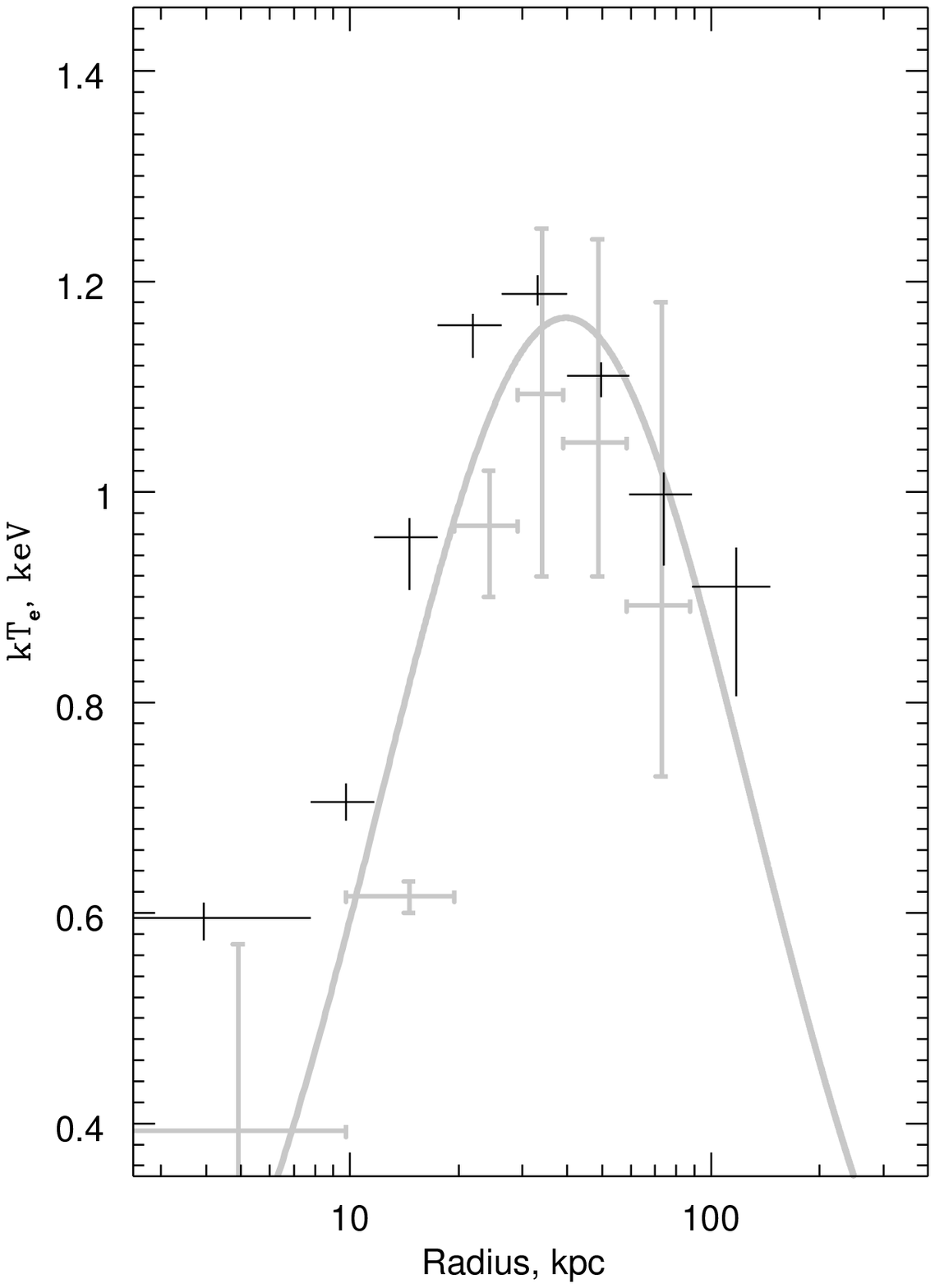}}

\figcaption{The radial temperature profile for NGC5846 is shown. The ROSAT
  PSPC determinations from each annuli are represented by grey crosses. ASCA
measurements are represented by black crosses. Errors are shown at the
1$\sigma$ confidence level. An analytical fit to the data points is shown by
a grey line.
\label{te_pro}}
\medskip

Figure \ref{te_pro} shows good agreement between the radial temperatures
derived from ASCA and those from ROSAT, particularly outside the cooling
region.  Cooling is important in the central 20~kpc.  Outside this region,
the coronae is approximately isothermal within 50~kpc. Outside 50~kpc, the
ASCA data suggest a decrease in temperature, which also is seen in the ROSAT
map of the projected temperature (Fig.\ref{te_map}). In Fig.\ref{te_pro} we
show a fit to our radial temperature measurements, which we use in Section
4.1 to derive the total mass profile.

\bigskip
\centerline{\includegraphics[width=3.25in]{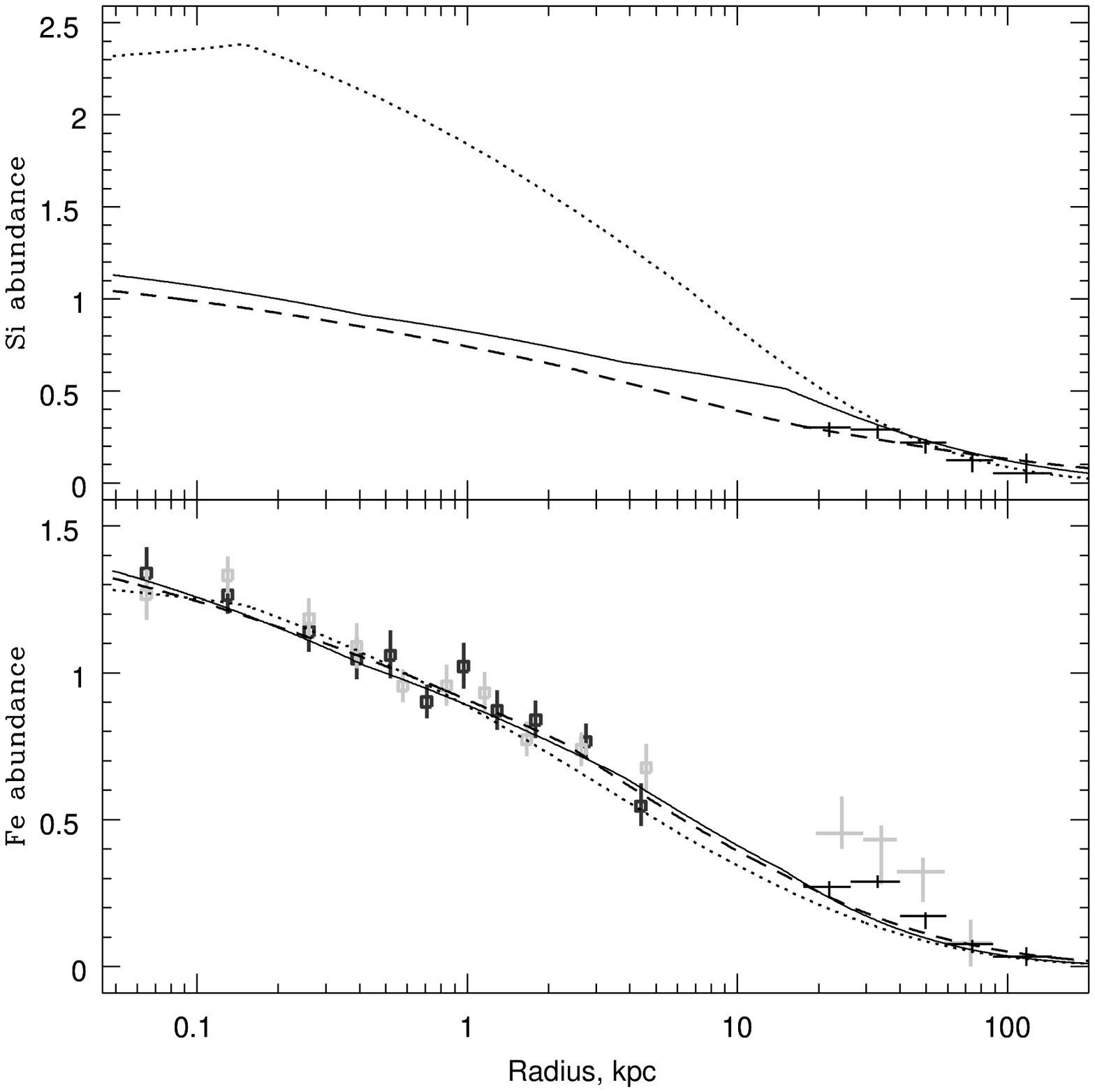}}

\figcaption{Radial profiles for the iron and silicon abundances observed in
  NGC5846.  The ROSAT PSPC points for iron are represented by grey crosses,
  while ASCA points are black. Errors are shown at the 1$\sigma$ level.
  Squares represent optical data on Fe, derived from FFI data for both the
  major (in grey) and minor (in black) axis of NGC5846.  (All iron
  abundances were converted into our adopted solar units of ``4.68e-5'' by
  number.) Black curves show the best-fit of the three models to the data
  with solid, dotted and dashed lines denoting models with Salpeter,
  Arimoto-Yoshii or Kroupa IMF's, respectively.
\label{ab_pro}}
\medskip

With the energy resolution of ASCA and ROSAT, we can derive the distribution
of iron and silicon, outside the central cooling region.  In Figure
\ref{ab_pro} we present the radial profiles for the abundances of Si and Fe,
derived from ASCA data (also given in Table 2), and the Fe profile, derived
from ROSAT data. Abundance units are as stated in Section \ref{sec:spe}.
Both Si and Fe profiles decrease with radius.  For comparison, in this
figure we also present the stellar Fe abundance, calculated from the Fisher,
Franx \& Illingworth (1995, hereafter FFI) data on Fe$_{4668}$ using the
modeling of line strength indices by Worthey (1994) for a galaxy age of
12~Gyr, chosen using the values H$_{\beta}=1.44$ and Fe$_{4668}=6.40$
from the FFI data on the galaxy nucleus. The ``age-metallicity degeneracy'',
as summarized by Kodama \& Arimoto (1997), has little effect on results for
giant ellipticals.  Also, there is good agreement in the simulations by
different researchers.


\begin{table*}

{\renewcommand{\arraystretch}{1.4} \renewcommand{\tabcolsep}{0.5cm}
\begin{center}
\tabcaption{\centerline{\footnotesize
Measurements of Si and Fe in NGC5846${^a}$.}}
\footnotesize

\begin{tabular}{cllll}
\hline
\hline
\vspace{1pc}
radii & Si & Fe & $M_{Si}$ & $M_{Fe}$\\
\hline
9.1--15. & 0.053 (0.00--0.16) & 0.034 (0.002--0.065) &8.18 (5.7--12.4)& 16.58 (12.6--19.8) \\
6.1--9.1 & 0.123 (0.06--0.16) & 0.076 (0.046--0.090) &6.10 (4.8--6.8) & 12.91 (11.1--13.7)\\ 
4.1--6.1 & 0.220 (0.16--0.24) & 0.171 (0.128--0.185) &4.04 (3.4--4.3) & 9.50  (8.3--10.0)\\
2.7--4.1 & 0.290 (0.24--0.31) & 0.288 (0.268--0.310) &2.45 (2.0--2.6) & 6.20  (5.4--6.5)\\
1.8--2.7 & 0.302 (0.25--0.33) & 0.271 (0.232--0.290) &1.19 (1.0--1.3) & 2.85  (2.6--3.0)\\
\hline
\end{tabular}
\end{center}
}
\vspace{1pc}

{$^a$}{\footnotesize ~ ASCA SIS results. Units for radii are arcminutes;
element abundances are given in solar units, as defined in the text;
cumulated masses are given in units of $10^6$ \msun\ and represent
integrated values. Si and Fe data are given as the best-fit values and the
corresponding 68\% confidence level intervals.}

\end{table*}

Although overall abundances of other elements, like Ne and Mg, could be
constrained from our ASCA analysis of NGC5846, the emission from these lines
is too weak to derive a radial distribution. Even placing respective upper
limits is not straightforward, due to the overlap of emission from these
elements with the poorly known Fe L-shell complex (e.g. Mushotzky \etal
1996). 

\bigskip
\centerline{\includegraphics[width=3.25in]{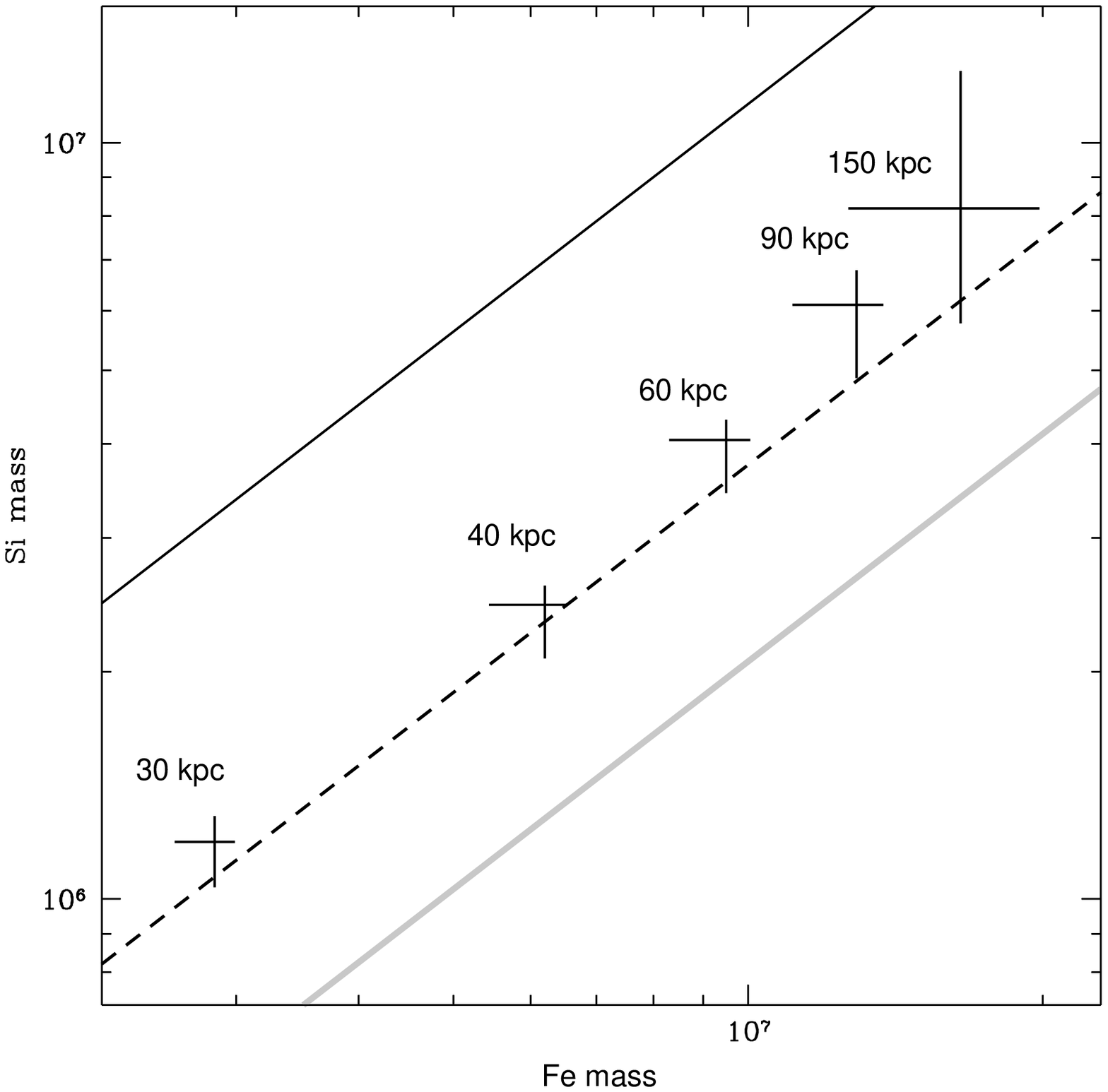}}

\figcaption{Observed Si mass vs Fe mass, integrated with radius. Crosses
  represent ASCA SIS determinations, omitting the data for the inner 20~kpc
(this omission results in somewhat lower values for element masses stated
for $<30$ kpc).  Every point corresponds to the mass within a certain
radius, also denoted on the figure.  The theoretical mass ratio for Si to Fe
from SN~Ia (TNH93) is shown as the bottom grey line. Dashed and solid black
lines represent a solar ratio of Si/Fe and three times solar, respectively.
\label{si_fe}}
\medskip

While the Fe abundance level provides constraints on the duration of star
formation and the SN~Ia rates, comparison of Si and Fe abundances determines
the role of different types of SNe in the chemical enrichment.  Such an
analysis for groups and clusters of galaxies was performed by Mushotzky
\etal (1996), Fukazawa \etal (1998) and Finoguenov \& Ponman (1998).  Clusters
reveal Si/Fe ratios favoring the prevalence of SNe~II in the overall
enrichment with Si/Fe increasing with cluster temperature (Fukazawa \etal
1998). On the other hand, groups and the centers of clusters are
characterized by a high input from SNe~Ia (Fukazawa \etal 1998, Finoguenov
\& Ponman 1998) with a significant dispersion in the Si/Fe ratio around the
solar value (Finoguenov \& Ponman 1998). The observed Si to Fe mass ratio
for NGC5846 is similar to that found for groups, although the total iron
mass converges to an IMLR (Iron Mass to Light Ratio, calculated as
$M_{Fe}$/$L_B$ in \msun/\lsun) of $0.9\pm0.2\times 10^{-4}$ (with $1\sigma$
uncertainty), which is much lower than in groups of galaxies (\eg\
Finoguenov \& Ponman 1998).  This evidence for substantial enrichment of an
elliptical galaxy by SNe Ia is an important indicator of post-formation
SN~Ia activity, since prevalence of SN~II in stellar chemical enrichment is
expected for systems with a short duration of star formation.

Analysis of the X-ray emission, in the 3--6 keV energy band, shows no
extended ``hard'' emission around NGC5846, in excess of that associated with
the 1 keV corona. As shown in Figure \ref{hard_pro}, the hard emission, in
excess of the best fit MEKAL model, is consistent with arising from an
unresolved source located at the galaxy center.  Nevertheless, following a
suggestion from the referee, we determined the limits on a hard component,
assuming a diffuse origin. For that, we have chosen one large region on the
detector, calculated and subtracted the contributions from all the ``soft''
components we have identified. The residual spectrum was fit in the 3--6 keV
energy band, assuming a bremsstrahlung spectrum of 5~keV temperature.  The
corresponding luminosity in the 0.5--4.5~keV band is
$3.6\pm0.7\times10^{40}$\ergss, although vignetting effects were not taken
into account, and could lead to an amplification of the measured flux by a
factor of 1.1 for a 1\amin\ source size up to a factor of 1.4 for a
10\amin\ size.  This value is comparable to the ``expected'' value of
$6.\times10^{40}$\ergss, taken from the work of Matsumoto \etal (1997,
Ma97). However the flux associated with this component can be explained by
the variations of the background, since the flux is only $\sim20$\% above
the calculated background level, compared to the expected 10\% variations
between the actual and predicted background fluxes. If a galactic population
contributes hard X-ray emission, that emission should be concentrated within
the confines of the optical galaxy.  For NGC5846, the observed hard emission
in the galaxy core corresponds to a luminosity of
$6.3\pm2.0\times10^{39}$\ergss.  The flux also could arise from a
low-luminosity AGN.

\medskip
\section{ NGC5850}
\medskip

As noted in section {\it 2.1.1}, NGC5850 (SBb galaxy, Sandage \& Tammann
1988, z=0.0084) is detected in both ROSAT and ASCA images. Although the
extended nature of the emission is evident in the ROSAT image, the
significance of the detection is insufficient to allow a separate spectral
analysis of the bulge and arm components. With ROSAT, using the MEKAL model,
we find $kT_e=0.32^{+0.41}_{-0.15}$ keV (the abundance is uncertain). The
spectral fit yields $\chi_r^2=0.80$ for 24 degrees of freedom.  We measure a
flux for NGC5850 from 0.5 to 2.0 keV of $5.06\times10^{-14}$\ergscmsec\
(which corresponds to a luminosity of $1.6\times10^{40}$\ergss, assuming a
distance to NGC5850 of 51 Mpc).  For the ASCA spectral analysis, we again
use the MEKAL model and also fit the contribution from NGC5846 to the region
around NGC5850. We find a gas temperature of $0.47^{+0.16}_{-0.11}$ keV,
with an upper limit on the heavy element abundance of Z/Z$_\odot<0.04$.
However, the quality of the fit is poor ($\chi_r^2=1.46$ for 47 d.o.f).  An
equally good fit is obtained for a power law spectrum with absorption. The
constraints on the spectral parameters are \nhl=1.2$^{+8}_{-1}\times
10^{21}$ cm$^{-2}$ with $\alpha>2.3$ ($\chi_r^2=0.88$ with 24 degrees of
freedom) and \nhl$<4\times 10^{21}$ cm$^{-2}$ with
$\alpha=4.1^{+2.6}_{-0.9}$ ($\chi_r^2=1.45$ for 47 d.o.f) for ROSAT and ASCA
data respectively. We note that the source spectrum is softer than for
low-luminosity AGN found in the centers of some spirals (Iyomoto \etal
1997), but comparable to the soft spectrum observed in the extended emission
for spirals (Marston \etal 1995; Read, Ponman and Strickland 1997).

\medskip
\section{ Discussion}
\medskip

\subsection{ Gravitating Mass Determination}

\bigskip
\centerline{\includegraphics[width=3.25in]{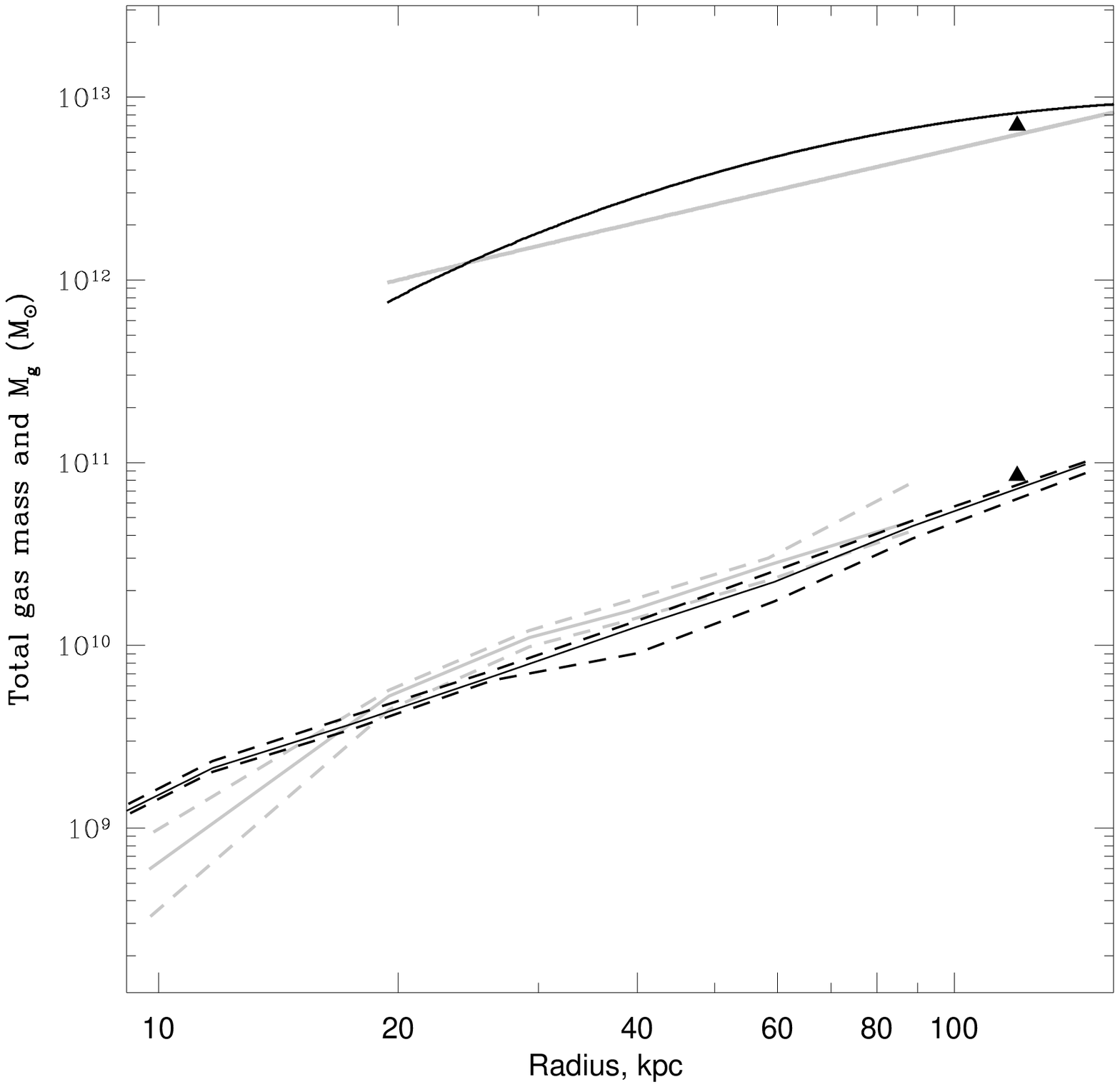}}

\figcaption{Total mass and gas mass profiles for NGC5846 as a function of
  radius. The gas mass determinations are shown as black curves and grey
  curves for ASCA and ROSAT data, respectively.  Dashed lines indicate the
  $3\sigma$ confidence intervals for these mass determinations.  The total
  mass profile is shown for comparison assuming an isothermal corona (grey
  line) and using the observed gas temperature profile (black line). Black
  triangles represent measurements from {\einstein} observations of the
  total mass and gas mass.
\label{mass_pro}}
\medskip

In Figure \ref{mass_pro} we show the gravitating mass profiles derived using
the equation of hydrostatic equilibrium and the data on temperature and
density distributions. We choose the parameters for the density
distribution, given in Table 1, corresponding to the analysis of the surface
brightness profile, which take into account the effects of temperature and
abundance changes with radius.  The temperature distribution was modeled in
two ways.  First we used an isothermal gas with a mean temperature 0.85 keV
(grey line). Second, we used the temperature variation from an analytic fit
to the ASCA and ROSAT data (black line) in the form $ A_3 e^{A_1 log(r)^2 +
A_2 log(r)} $, chosen to provide an easy way to calculate the $ d log(T) / d
log(r) $ term in the mass determination. The maximum deviation in the mass
between these two approaches is less then $\pm30$\% at any radius.
Nevertheless, with the decrease in gas temperature beyond 5\amin\ (50 kpc)
suggested from the ROSAT and ASCA data, the mass profile converges to a
M/$L_B$ value of $53\pm5$ \msun/\lsun\ at a radius of 230~kpc (note that the
temperature measurements extend only to 150~kpc). The total mass determined
from {\einstein} data (Biermann \etal 1991) is shown with a triangle and is
in good agreement with our determinations. In Figure \ref{mass_pro} we also
show the gas mass derived from spectral analysis of the ROSAT (grey lines)
and ASCA (solid lines) data. The gas mass is typically only a few percent of
the total mass, similar to that found in other elliptical galaxies (\eg\
Forman, Jones \& Tucker 1985).

\subsection{ Stellar metallicities, SN Ia rates and
questions of star formation.}

It is generally agreed that early type galaxies are currently passively
evolving stellar systems, where significant star formation was cut off at
early epochs by a galactic wind (\eg\ Ciotti 1991, David \etal 1991). The
large amount of hot gas presently found in E's is attributed to stellar mass
loss, and as such, should be characterized by stellar metallicity and
stellar velocity dispersion, with supernovae supplying additional elements
and energy into the interstellar medium. Previous work, where this scheme
was adopted, met a mismatch between stellar abundances, derived via modeling
of optical measurements, and the tremendously low metallicities found in the
X-ray gas, which imply both a low metal content in stars and a SN~Ia rate
which is lower then measured in optical searches (\eg\ Turatto \etal 1994).

Perhaps the first step towards resolving this apparent discrepancy was the
discovery of strong abundance gradients in the stellar content of
ellipticals (\eg\ Carollo, Danziger and Buson 1993, hereafter CDB; FFI).
Combining these results with the presence of gas inflow, LM were able to
explain the low X-ray abundances, although the problem of low SN~Ia rate
persisted.

Triggered by numerous exciting achievements on the observational side, a
number of theoretical models were developed to explain different aspects of
the chemical evolution in ellipticals (\eg\ David \etal 1991). However,
detailed modeling of the radial behavior of the chemical content of the
X-ray gas in elliptical galaxies has yet to be done. Nevertheless, there are
two cornerstones that define this modeling. First is a set of models of
elliptical galaxies with different mass, calculated by Matteucci (1994) and
second is the suggestion by CDB that the optical abundance gradients result
from the duration of star formation, which is shorter at larger radii, once
a standard dissipative collapse model (Carlberg 1984) is adopted for the
formation of ellipticals.

Assuming that NGC5846 is typical, we can use our measured profiles of 
silicon (an $\alpha$-element) and iron to constrain the enrichment models
for gas-rich ellipticals. We build our model using the results of a
classical wind model, presented in Table 1 of Matteucci \& Gibson (1995,
hereinafter MG) for three choices of the IMF, namely Salpeter (S), Arimoto
\& Yoshii (AY) (with A=0.02) and Kroupa \etal (K). The models are
constrained to reproduce the optical stellar abundance measurements as well
as the X-ray determined abundance gradients at radii exceeding 2\amin, where
effects of a central cooling flow are negligible.  We adopt the time of the
galactic wind onset as a measure of the duration of star formation. 

In detail, the modeling consists of adjusting the onset of the galactic wind
to fit the derived stellar Fe and X-ray Si and Fe abundance measurements at
all radii, using the predictions from one-zone models of MG.  Enrichment
from SN~Ia is not considered, due to its present uncertainty, and X-ray
abundances are assumed to be produced by stars, although we will later
return to the predictions for SN~Ia rate implied by the model.  While a
complete modeling including the gas mass exchanges, such as the one
presented in Martinelli \etal (1997), is beyond the scope of this paper, we
will be able to draw a number of important conclusions from our simple
model.

We parameterize the onset of the galactic wind as
\begin{equation}
t_{gw} = t_{gw_o} \; exp({-({r / r_e})^{-\gamma}})   
\end{equation}

where $r_e$ is fixed to 6.3\asec\ (1/10 of the effective radius for
NGC5846), and $t_{gw_o}$ and $\gamma$ are allowed to vary to match the
actual abundance steepening.

\begin{minipage}[H]{8.5cm}
\begin{table}[H]

{\renewcommand{\arraystretch}{1.4} \renewcommand{\tabcolsep}{0.5cm}
\begin{center}
\tabcaption{\centerline{\footnotesize
Duration of star formation in NGC5846${^a}$.}}
\footnotesize

\begin{tabular}{cll}
\hline
\hline
\vspace{1pc}
IMF & $t_{gw_o} (Gyr)$ & $\gamma$\\
\hline
S  & 2.77 (2.67--2.87) & 0.359 (0.35--0.37)  \\
AY & 0.86 (0.82--0.90) & 0.414 (0.40--0.43)  \\
K  & 2.51 (2.42--2.53) & 0.301 (0.29--0.31)  \\
\hline
\end{tabular}
\end{center}
}
\vspace{1pc}

{$^a$}{\footnotesize ~ Results of fitting the NGC5846 abundance
profile with chemical enrichment models from Matteucci \& Gibson (1995).
Errors are given at 90\% confidence level for one parameter of interest. See
text for details of analysis.}

\vspace{1pc}

\end{table}

\end{minipage}

The results of fitting the measured abundance profiles for NGC5846 with the
three models including different IMFs are presented in Table 3.  The
best-fit curves for the predicted stellar abundances of Si and Fe are
plotted in Fig.\ref{ab_pro}. From the slopes considered for the Initial Mass
Function, the Kroupa IMF provides the best description of the Si data, while
all the models provide only an approximate fit to the X-ray data on Fe. Yet,
given the complexity with the modeling of SN~Ia rate discussed below, we
consider the results of this fitting as indicative, rather then an evidence
for any particular choice of IMF slope.

Figure \ref{mod_tgw} suggests that a strongly decreasing duration of star
formation with distance from the center of the galaxy causes the decline in
stellar metallicity and SN~Ia rate with radius. These results suggest that
the duration of star formation in the outskirts of NGC5846 is extremely low,
compared to the central value. At a radius of 100 kpc, the star formation
duration would be only about $10^7$~yr. However, with such a short period of
star formation, one should observe a SN~II like element ratio, which is
contrary to our measurements for Si and Fe, although a drastic difference
between models and X-ray data in Fig.\ref{ab_pro} is seen mostly for X-ray
data inside 60 kpc. While below, we consider a dependence of SN~Ia rate on
the metallicity of the progenitor star, an alternative way to explain both
the low Si/Fe ratio and low Fe abundance is to introduce a prolonged, but
inefficient star formation. We propose a scenario in which the declining
metallicity results from a change in the efficiency of star formation with
galaxy radius, while the duration of star formation does not vary
significantly with radius. The constancy in the duration of star formation
can be explained by the short time required for a galactic wind to terminate
star formation everywhere in the galaxy.  A small radial dependence in the
time for star formation triggering is negligible compared to the typical
duration of star formation $\sim3$~Gyrs, required for the SN~Ia to play a
significant role in the enrichment (Yoshii \etal 1996).

\bigskip
\centerline{\includegraphics[width=3.25in]{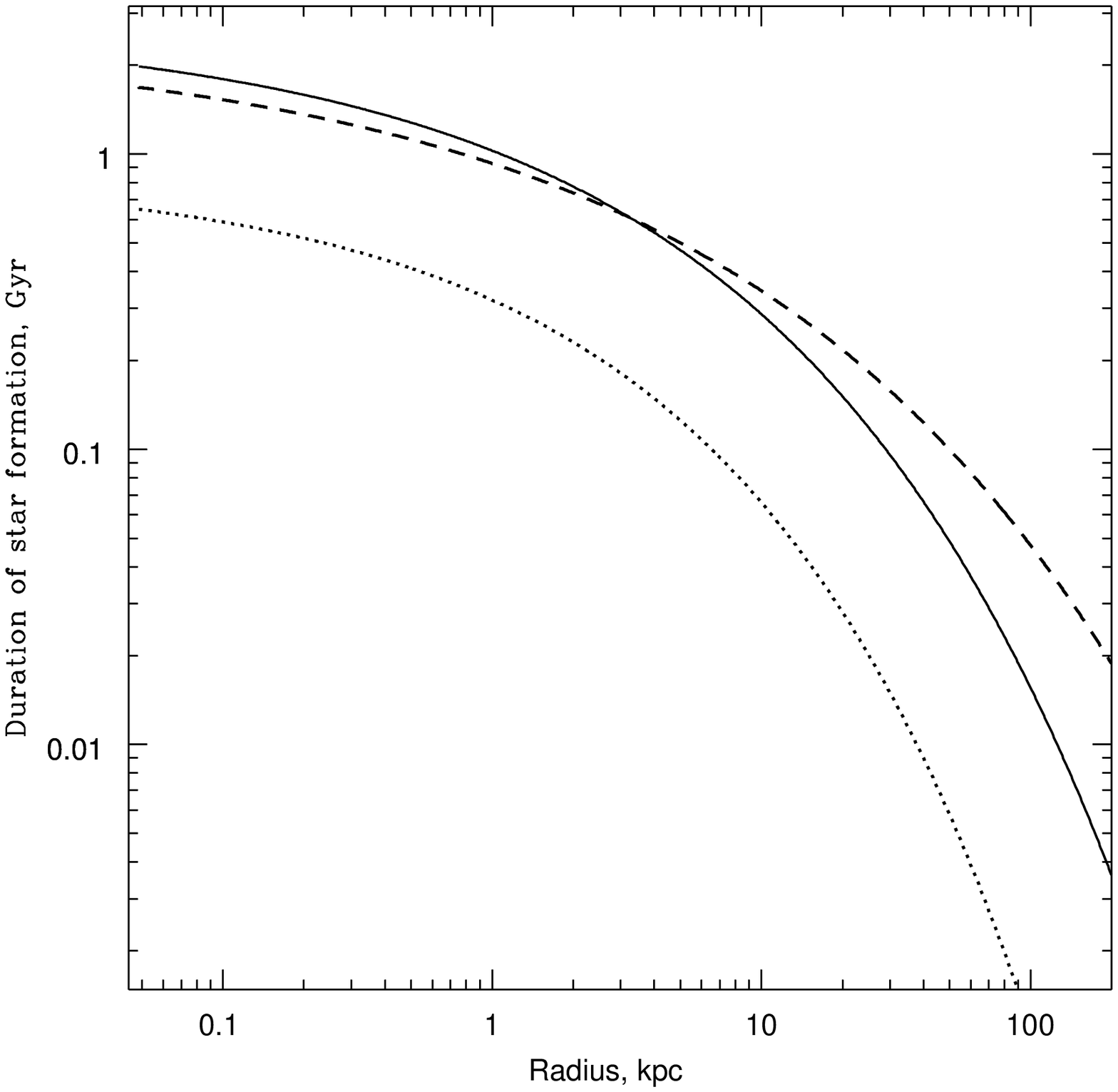}}

\figcaption{Duration of star formation as a function of galaxy radius,
  implied by fitting the X-ray and optical abundance measurements with a set
  of theoretical models (see the text for details). The curves are coded as
  follows: solid, dotted, and dashed lines denote respectively a choice of
  Salpeter, Arimoto-Yoshii, or Kroupa IMF.
\label{mod_tgw}}
\medskip

A drop in the efficiency of star formation with radius could result from a
corresponding decrease in the density of gas clouds. This scenario would
result in lower metallicities at larger radii, but similar abundance ratios
throughout the galaxy, as we find for NCG5846. The observation that the
abundance ratio of Mg to Fe does not change within an individual galaxy, but
changes quite drastically from galaxy to galaxy (Worthey \etal 1992) also
supports this scenario. In addition, since this scenario just modifies the
standard modeling by including an additional parameter, the efficiency of
star formation as a function of radius, it should also meet other
observational restrictions discussed in Matteucci (1994).

Considering the predictions for present-day SN~Ia rates, illustrated in Fig.
\ref{mod_snr}, we note that for all the models, the resulting SN~Ia rate in
SNU is zero, except in the central region ($r<20$~kpc) of NGC5846. This
result is quite different from that found, when a constant star formation
rate with radius is assumed (as \eg\ in LM). The behavior of the SN~Ia rates
in MG probably results from the choice of the evolutionary tracks from Van
den Bergh \& Bell (1985).  As was shown by Bazan \& Mathews (1990), these
tracks correspond to shorter main sequence lifetimes for low metallicity
stars due to their higher luminosities. In calculating the theoretical SN~Ia
rates, normalized to the blue luminosity, a low metal content in stars
implies both {\it faster} evolution (and as a consequence a stronger
decrease with time) of the SN~Ia rate and enhanced stellar luminosities. We
have verified this suggestion, by a detailed modeling of the SN~Ia rate, as
described in David \etal (1991 and references therein), and found that the
influence of the parameters described above results in suppressing the
present SN~Ia rate for low Z stars by $\sim 20$ times.

\bigskip
\centerline{\includegraphics[width=3.25in]{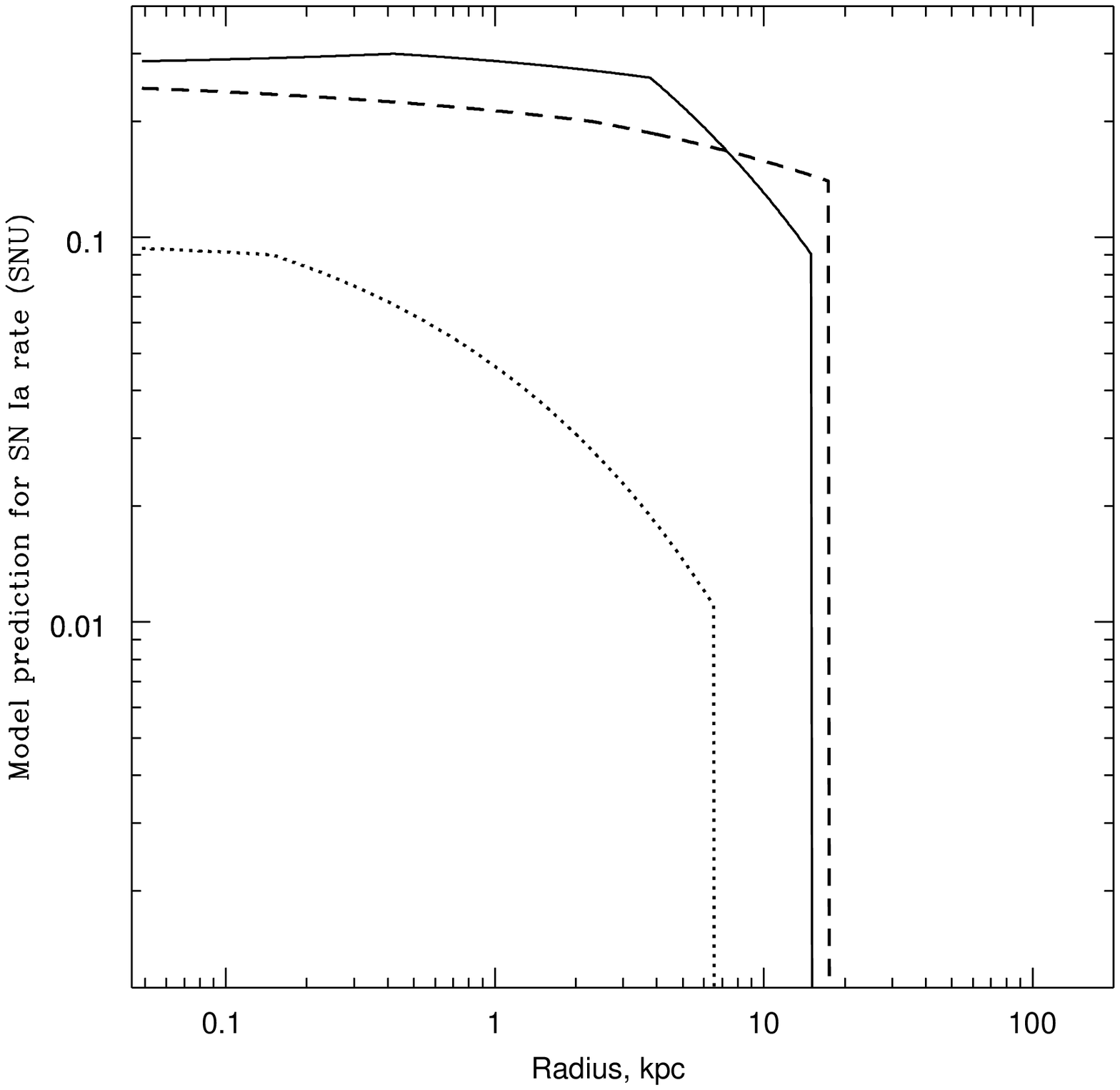}}

\figcaption{The present-day SN~Ia rate as a function of radius, implied by
  fitting the X-ray and optical data with a set of variable star
formation models (see the text for details).  Coding of the curves is the
same as in Fig.\ref{mod_tgw}. All of these show a sharp decline at large
radii in the SN~Ia rates.
\label{mod_snr}}
\medskip

Nevertheless, since the evolution of the stellar mass loss is based on the
lifetime of stars on the main sequence, mass loss for the low metallicity
stars at outskirts of the galaxy is also lower, and the ratio of SN~Ia over
stellar mass loss does not change significantly in our modeling, compared to
estimates from LM.  Therefore, the strict X-ray limits, usually calculated
for the solar metal composition, are still a problem for the rate of SN~Ia
events found in optical searches (for a recent review of this issue, see
Arimoto \etal 1997). However, the X-ray limits on the SN~Ia rate correspond
to a different companion mass for the SN~Ia progenitor. Thus one can, in
principle, by changing the distribution of the probability of an SN~Ia event
as a function of mass ratio in a binary, satisfy both the X-ray and optical
requirements.  Another possibility, discussed in Timmes, Woosley \& Weaver
(1995) and Kobayashi \etal (1998) is a dependence of the SN~Ia rate on the
metallicity of the progenitor star.  In this respect, we would like to point
out, that if a dependence of the SN~Ia rate on stellar metallicity is
assumed, the stellar Si/Fe ratio could be larger than that observed for the
hot gas, altering our conclusions on the efficiency of star formation.

To illustrate the problem with SN~Ia rates, we calculate the upper limits on
the SN~Ia rate, assuming that all the metals in the outskirts of NGC5846,
found in our X-ray measurements, are due to SN~Ia explosions. We then use all
the specific values, calculated for the solar metal composition, so these
limits can be directly compared to the SN~Ia rate predictions for the galaxy
center.

Following LM, the X-ray derived abundances can be expressed as

\begin{equation} \label{fe_eq}
[Z/H]=[Z/H]_* + {f_{\rm SN} \; {\rm SNU} \; M_{{\rm
      SN},_{Z}} \over {M_* / L_B} \; \; {\dot{M}_* / M_*}}\;,
\end{equation}

where $f_{\rm SN}$ is the SN~Ia rate in units of an SNU (1 SNU = one event
per 100 yrs ${ L_B}$/10$^{10}$), equation has $M_{{\rm SN},_{Z}}$, but here
is uses $M_{{\rm SN},_{Fe}}$ -- the mass of iron released in each SN~Ia
event, and ${\dot{M}_* / M_*}$ is the stellar mass loss,
which we adopt as 3 (for the Salpeter and Kroupa IMFs) or 5 (for the Arimoto
\& Yoshii IMF) $\times10^{-20}$ sec$^{-1}$ (following calculations by
Mathews 1989). In such an approximation, we implicitly assume that the SN~Ia
rate changes with time, similar to the stellar mass loss.  We adopt a
central $M_*/L_B=11.3$\msun/\lsun\ by scaling the value of 19 from Lauer
(1985), obtained for $H_o=84$km sec$^{-1}$ Mpc$^{-1}$. Z denotes the metal
considered. In placing upper limits on the SN~Ia rates, we neglect the
stellar (first) term on the right side of the equation.

The upper limit on the SN~Ia rate using our measurements for silicon in
the 5\amin--15\amin\ annulus is $0.012$ SNU ( 1 SNU = 1 event (100
yr)$^{-1}$ ${L_B}$/10$^{10}$), for a Salpeter IMF (1.35), and 0.021 SNU, for
an IMF slope of 0.7. This calculation is independent of the modeling for the
L-shell emission for iron and the effects of a central cooling flow or
incomplete thermalization attributed to the central regions.  Corresponding
limits, using Fe, are 0.006 and 0.010 SNU for IMF slopes of 1.35 and 0.7,
respectively.

The possibility of significant amounts of additional iron (or silicon) in
the form of dust is low, since the dust mass inferred from recent
observations is $\sim 7\times 10^3$\msun\ (Goudfrooij \& Trinchieri 1998).
While dust is not likely to play an important role in the abundance, it
could be a major factor for the energy balance of the hot gas in the center
of NGC5846 (Goudfrooij \& Trinchieri 1998).

Recently (\eg\ Matsushita \etal 1998), the X-ray emission associated with
some ellipticals was suggested to have a group component, based on a
``beta-model'' analysis of the surface brightness profile.  While our
chemical enrichment modeling is insensitive to the presence of the
surrounding group potential (LM), it is sensitive to ascribing the detected
metals and gas to being a product of stars at the corresponding radius.
Alternatively, detected elements could be treated as having been expelled
from the galaxy center and held in the group potential. The observed X-ray
gas would consist of the galaxy products: SN~Ia output and stellar mass loss
{\it recently mixed} with the intragroup gas, which should have low heavy
element abundances, due only to the contribution of other galaxies in the
group.  The abundance gradient indicates the degree of mixing, while the
constancy of the abundance ratio with radius, measured in X-rays, results
because the SN~Ia rate evolves like the stellar mass loss (\eg LM).

However, this scenario contradicts the observed picture for NGC5846.  The
amount of iron should indicate the time of group formation. For the SN~Ia
rate at the galaxy center of 0.16 SNU, implied by optical observations
(Turatto \etal 1994), the observed iron mass in NGC5846 would be reproduced
in less then 1 Gyr. Thus the group formation must be recent, which is
surprising, considering the hierarchical clustering hypothesis. A possible
solution would be if the group formed long ago, but the energy released by
SN~Ia continued to drive a galactic wind until the recent epoch. The
calculation shows that the energy associated with the release of iron via
SN~Ia explosions is less then 10\% of the thermal energy of the gas at any
radius in NGC5846. It is dubious that such a small contribution recently
played a dramatic role.

Finally, we comment on the energy balance in NGC5846. The kinetic energy of
the stars is only 1/3 the observed thermal energy of the X-ray gas.  The
rest of the energy may have come from the gravitational infall of the gas.
We can use our measured mass profile to verify this possibility.  We find
that sufficient energy can be produced if the gas falls in adiabatically
from only 1/3 of the initial radius.  An alternative explanation for the
energy budget of the X-ray emitting ellipticals, was proposed by Mathews \&
Brighenti (1997), in which the temperature of the gas is due to the inflow
of hot circumgalactic gas. We note, that a decline of the temperature with
outer radius in NGC5846 contradicts the basic assumptions of this scenario,
since it implies that the circumgalactic gas, if it exists in a considerable
amount, is too ``cold'' to reproduce the measured temperature profile by
simple gas mixing with ``cold'' gas of the galaxy, assumed in the model.
Note, however, that in NGC5044 and NGC4472, such a drastic temperature drop
is not observed (David \etal 1994; Finoguenov \& Ponman 1998; Forman \etal
1993).

\section{ Conclusions}

Applying methods of spatially resolved spectroscopy, we studied X-ray
emission from hot gas in the elliptical galaxy NGC5846. We detect cooler gas
both toward the galaxy center and in the outer parts of the galaxy. Such a
decreasing temperature profile, if continued to larger radii, leads to a
convergence of the total mass in NGC5846 to $9.6\pm1.0 \times
10^{12}${\msun} at a radius of $\sim$230~kpc.

Our study of the heavy element distribution reveals radial abundance
gradients in both Fe and Si, but a constant Si/Fe ratio at the level
favoring a greater retention of SN~Ia products, compared to SN~II.
Comparison of optical and X-ray data shows remarkable agreement in the
radial distribution, with Fe abundance falling from a value of 1.3 times
solar at the center to a value of 0.1 solar at 100~kpc. We apply the
theoretical models of Matteucci \& Gibson (1995) to quantify our data. These
models are characterized by a decrease in the duration of star formation
from a few Gyr in the central region of the galaxy to a duration of only
$\sim0.01$ Gyr at 100~kpc. Alternatively, the decline in metallicity with
galaxy radius may be caused by a variable efficiency of star formation as
a function of the galactocentric distance.

The model prediction for the present-day SN~Ia rate is non-zero only in the
galaxy center and is caused by faster evolution of the low-metallicity stars
found at the galaxy outskirts. Nevertheless, our calculation shows that
tight X-ray limits, usually calculated for the solar element composition,
remain a problem, since the SN~Ia rate evolves like the stellar mass loss.
To provide an acceptable explanation of X-ray and optical measurements, we
suggest an introduction of a dependence of SN~Ia rate on the progenitor star
metallicity, \eg\ as is suggested by Kobayashi \etal (1998).

We also investigated alternative mechanisms for the low iron abundance in
NGC5846.  One alternative explanation for the low metallicity is that the
primary emission arises from a group and NGC5846 only pollutes the
surrounding intragroup media. However, under the assumption of standard
SN~Ia rates, the low iron mass detected leads to a recent time for group
formation ($<1$ Gyr), which contradicts this scenario. Expelling the iron
from the gravitational potential of NGC5846 via recent galactic winds,
driven by SN~Ia's, also is unlikely, since the energy release by SN~Ia's is
small, less then 10\% of the thermal energy of the gas.

\vspace{0.5cm}

This research has made use of data obtained through the HEASARC Online
Service, provided by the NASA/GSFC.  C.J. and W.F. acknowledge support for
this research from the Smithsonian Institution and the AXAF Science Center
NASA Contract 16797840. A.F. was supported by the Smithsonian predoctoral
fellowship program. Authors wish to thank Maxim Markevitch for useful
discussions and the referee for helpful comments.

\appendix 
\section*{Incorporation of the ASCA Point Spread
Function into Spectral Analyses}

In this chapter we will explore the effect produced by the broad energy
dependent ASCA PSF on study of diffuse objects.

A general expression for the observed spectrum reads
\begin{equation}
 \eta(y,e^\prime) = \int_{E} r(e^\prime \mid e) 
\beta(y) \int_{X} p\tild(y \mid x, e) \nu\tild(x,e) dx de 
\end{equation}

\noindent
with the following definitions:

\noindent
$\nu\tild(x,e)$ -- emission from the source, where x stands for spatial
coordinates and $e$ -- energy 

\noindent
$p(y \mid x, e)$ -- the Point Spread Function
including telescope effective area 

\noindent
$y$ -- refers to projected coordinates on the detector plane, 

\noindent
$r(e^\prime \mid e)$ -- is the detector
response matrix yielding the probability distribution of detecting a
photon of energy $e$

\noindent
$\beta(y)$ = 0 for ``not valid'' detector pixels like
gaps or chips not used in the observation (different CCD clock mode), 1 -
otherwise.  

\noindent
$\eta(y,e^\prime)$ -- is the observed photon spectrum in units
of ph sec$^{-1}$ cm$^{-2}$ keV$^{-1}$ arcmin$^{-2}$

\noindent
where  accumulated spectra are:
\begin{equation}
 \eta_j(e^\prime) = \int_{Y_j} \eta(y,e^\prime) dy 
\end{equation}
and $Y_j$ is a set of spatial regions in the image data, such that \ \ \
\(Y_{j_n}\bigcap Y_{j_m} =\emptyset \) \ for any \(j_n \neq j_m\). This is
required for further use of standard likelihood estimators.

A {\it key assumption} is that spatial and spectral properties can be
separated

\begin{equation}
 \nu\tild(x,e) = \sum_i { \lambda_i(x)\nu_i(e)} \theta(x,X_i)\ \ 
,\ \ \ \ \ \int_{X_i} \lambda_i(x) dx = 1 
\end{equation}
where $\theta(x,X_i)=1$ if $x\in X_i$ and 0 otherwise, and $X_i$ is a set of
spatial regions in the object image with \ \ \ \(\bigcup_i X_i = X \) \ \ \
by definition. $\lambda_i(x)$ is the spatial distribution which we
take from the surface brightness measurements of the ROSAT PSPC.

Now, we can separate the spectral and imaging parts:

\begin{equation}
 \eta_j(e^\prime) = \int_{E} r(e^\prime \mid e) \sum_i \nu_i(e)
\int_{Y_j} \beta(y)\int_{X_i} p\tild(y \mid x, e) \lambda_i(x) dx dy de 
\end{equation}

The imaging part can be rewritten in a matrix form which further accounts
for different telescope pointings:\\
\begin{equation}
 a_{ij}\tild(e) = \int_{Y_j} \beta(y)\int_{X_i} p\tild(y \mid x, e)
\lambda_i(x) dx dy \ \ ,\ \ \ \ \  a_{ij}\tild(e)\equiv a_{ij}\tild(e,k) 
\end{equation}
 
\noindent
where $a_{ij}\tild(e,k)$ is a spatial correlation matrix, $k$ represents
different telescope pointings.

We define
\begin{equation}
 A_{ij}(e) = \sum_k t_k a_{ij}\tild(e,k) 
\end{equation}
where $t_k$ is the time spent at each pointing. With this definition
equation 5) becomes

\begin{equation}
 \eta_j(e^\prime) = \int_{E} r(e^\prime \mid e) \sum_i \nu_i(e) A_{ij}(e)
\ de\end{equation}

The procedure of fitting the data with spectral programs like XSPEC is as
follows:

\noindent
1) produce a set of models $ \nu_i(e)$ \\ 
2) mix them using calculated $A_{ij}(e)$  \\
3) convolve them with response matrix $r(e^\prime \mid e)$.

In the analysis presented in this paper, we used ASCA XRT ray-tracing
azimuthly averaged PSF for the calculation of the scattering from the central
part of the source emission (inside 5\amin\ radius) and the PSF data from
ASCA GIS measurements (Takahashi \etal, 1995) for the rest.

\end{document}